\documentclass[twocolumn,prd,showpacs,showkeys]{revtex4-1}
\usepackage{amsmath,amssymb,pgf,graphicx,color,url,array}
\usepackage{tabularx}
\usepackage{wasysym}
\usepackage{longtable}
\usepackage{ulem}
\newcommand{\be}{\begin{equation}}
\newcommand{\ee}{\end{equation}}

\begin{document}
\begin{flushright}
INR-TH/2015-017
\end{flushright}

\vskip -0.9cm
%--------------------------------------
\title{Probing Milky Way's hot gas halo density distribution using
  the dispersion measure of pulsars}

\author{Emin Ya. Nugaev}
\email{emin@ms2.inr.ac.ru}
\affiliation{Institute for Nuclear Research of the Russian Academy of Sciences, 117312, Moscow, Russia}
\author{Grigory I. Rubtsov}
\email{grisha@ms2.inr.ac.ru}
\affiliation{Institute for Nuclear Research of the Russian Academy of Sciences, 117312, Moscow, Russia}
\author{Yana V. Zhezher}
\email{zhezher.yana@physics.msu.ru}
\affiliation{Faculty of Physics, M.V. Lomonosov Moscow State University, 119991, Moscow, Russia}
\affiliation{Institute for Nuclear Research of the Russian Academy of Sciences, 117312, Moscow, Russia}
	%{\small $^{1}$Institute for Nuclear Research of the Russian Academy of Sciences, 117312, Moscow, Russia}\\
	%{\small $^{2}$Faculty of Physics, M.V. Lomonosov Moscow State University, 119991, Moscow, Russia}

%\author{G.~I.~Rubtsov}
%\email{e-mail}
%\author{S.~V.~Troitsky, Y.~V.~Zhezher}
%\email{e-mail}
%\affiliation{Institute  for Nuclear Research of the Russian Academy of Sciences,
%Moscow 117312, Russia}

\begin{abstract}
A number of recent studies indicates a significant amount of
ionized gas in a form of the hot gas halo around the Milky Way. The
halo extends over the region of 100 kpc and may be acountable for the
missing baryon mass. In this paper we calculate the contribution of
the proposed halo to the dispersion measure (DM) of the pulsars. The
Navarro, Frenk \& White (NFW), Maller \& Bullock (MB) and Feldmann,
Hooper \& Gnedin (FHG) density distibutions are considered for the gas
halo. The data set includes pulsars with the distance known
independently from the DM, e.g. pulsars in globular clusters, LMC, SMC
and pulsars with known parallax. The results exclude the NFW
distribution for the hot gas, while the more realistic MB and FHG
models are compatible with the observed dispersion measure.
\end{abstract}
\keywords{missing baryons -- hot gas -- pulsars -- dispersion measure}
\maketitle

\section{Introduction}
\label{sec:intro}

``Missing baryons'' is one of the long-standing problems of modern
astrophysics. The total baryon mass obtained from the cosmic microwave
background anisotropy measurements and from the data on the big-bang
nucleosynthesys exceeds the mass of the stars, interstellar and
intergalactic media~\cite{Nicastro}. The ratio of the baryon mass to
the total mass at high redshifts is estimated with better than 5\%
accuracy from the Planck data~\cite{Planck}. On the other hand we observe
less than the half of these baryons at present.

One of the possible solutions for this problem is that ``missing''
baryons reside in intergalactic medium (IGM). It is possible that the
haloes of the galaxies may be the dominant component of
IGM~\cite{Meiksin}. In this scenario the Milky Way's halo is a huge
bubble of gas with the radius up to 250 kpc~\cite{Miller} which may be
produced by the matter accretion from the satellite galaxies and intergalactic medium~\cite{Bell},~\cite{Lucia}.

Milky Way's gas halo may be considered as a composition of three general components:
cold gas, warm ionized medium (WIM) and hot ionized medium (HIM). First one is neutral
hydrogen, which extends for approximately 1.6 kiloparsecs into the halo region with mass
${3.2}^{+1.0}_{-0.9} \times 10^8 M_{\astrosun}$ \cite{Marasco}.

The WIM component is produced by strong UV radiation from O and B
stars. It mainly consists of the ionized hydrogen ($H_{II}$ regions)
and additionally of sulfur, nitrogen and oxygen ($S_{II}$, $N_{II}$,
$O_{I}$ - $O_{III}$ lines) \cite{Ferriere}, \cite{Haffner}. WIM
contains 30 \% of interstellar medium mass and its temperature is up
to $10^4$ K. The size of the warm gas halo is about 2 kpc which is similar
to the size of the neutral hydrogen component.

The third component is a hot ionized gas with the temperature of
$10^6$~K. It was first suggested by Spitzer in 1956~\cite{Spitzer} and
twenty years later proved by observation of UV absorbtion lines
\cite{York}, but it was still hard to determine the exact composition
and structure. The HIM is considered as one of the candidates for the
source of all ``missing'' baryons.  An interest for this solution has
grown in 2012 after the work of A.~Gupta and S.~Mathur
\cite{Gupta}. Analyzing the data from Chandra X-ray Observatory they
examined emission and absorbtion lines of active galactic nuclei
and have found a huge reservoir of ionized gas traced by $O_{VII}$ and
$O_{VIII}$ absorption lines extending to more than 139 kpc. The mass of the hot gas halo is
compatible with the ``missing'' mass and this may be considered as a
solution of the problem.

In this paper we study the contribution of the hot has halo proposed
in~\cite{Gupta} to the dispersion measure of the pulsars. The DM is
calculated for the case of the three models of halo density
distribution and compared to experimental data.

The present paper is organized as follows. First, in Section
\ref{sec:methods} we define a dispersion measure and overview the
Navarro, Frenk \& White (NFW), Maller \& Bullock (MB), and Feldmann,
Hooper \& Gnedin (FHG) density distribution models.  Next, in Section
\ref{sec:data} the data set of the pulsars with the distance known
independently of the DM is formed. The comparison of the data with the
calculations is given in Section \ref{sec:results}.

\section{Methods}
\label{sec:methods}

Electromagnetic waves emitted by a pulsar propagate through the
interstellar plasma. The plasma is dispersive, i.e. the
lower-frequency radio waves travel slower than the higher-frequency
ones. Propagation delay may be calculated as follows~\cite{Tanenbaum,Taylor}:
\begin{equation}
\delta t = \frac{e^2}{2 \pi m c {\nu}^2} DM,
\end{equation}
\noindent where $DM$ is defined as an integral of the electron density
along the propagation line:
\begin{equation}
\label{DM}
DM = \int_0^d n_e ds.
\end{equation}

Observing the pulsar at multiple radio bands one may measure the time
shift between the pulse profiles at these bands with the high
accuracy. This gives an instrument to probe free electron distribution
in the interstellar medium (ISM).

The calculations of the DM are based on the model of electron
distribution in the Milky Way's halo. There are two general models
conventionally used in this case \cite{Fang}: Navarro, Frenk \& White
and Maller \& Bullock.

Navarro, Frenk \& White model describes the dark matter mass distribution,
derived by simulating N-body system in the Standard Cosmological Model
\cite{Navarro}. It approximates equilibrium distribution of dark
matter for non-colliding particles:
\begin{equation}
\rho_g^{NFW} \left( r \right) = \frac{\rho_g}{x {\left( 1 + x \right)}^2}\,,
\end{equation}
\noindent where $r$ is the distance from Galactic center, $x = r / R_s$, $ R_s = R_v / C
$, where $R_v = 260\ \mbox{kpc}$ is virial radius associated with Milky Way's dark matter halo, $C$ is halo concentration.

The two particular cases are considered: density profile may mimic
dark matter distribution with its concentration $C_v$, and $C = C_v = 12$, or it obeys
low-concentration scenario with $C = 3$. We note, however, that the gas in
haloes has non-gravitating mechanism of formation and the density
profile for the dark matter is not directly applicable.

The Maller \& Bullock model was proposed in 2004~\cite{Maller}. It
describes a gas in hydrostatic equilibrium within the dark matter
halo. If hot gas doesn't radiate significantly, it obeys an adiabatic
law with polytropic index $5/3$ within the cooling radius $R_c$, $p
\varpropto \rho_h^{5/3}$. The dark matter is assumed to follow NFW
profile with the $C_v = 12$. The hot gas density
  distribution is the following:

\begin{equation}
\rho^{MB}_g \left( r \right) =
\begin{cases}
\varkappa_1 {\left[ 1 + \frac{3.7}{x} \ln{\left(1 + x\right) -
      \frac{3.7}{C_c} \ln{\left(1 + C_c \right)}} \right] }^{3/2},
\\ \hfill r < R_c,\\
\varkappa_2\ /\ r^2, \hfill R_c < r < R_v\,.
\end{cases}
\end{equation}

The parameter $C_c = R_c / R_s$ takes the value $C_c = 7$ according
to~\cite{Maller}.

 We obtain virial density ${\rho}_v$ for NFW distribution and
  constants $\varkappa_1$, $\varkappa_2$ for MB distribution from
the assumption that a sphere with its center in the Galactic center
and radius $R = 260\,\mbox{kpc}$ contains all the Galaxy ``missing''
mass:
\begin{equation}
4 \pi \int_0^{R} \rho r^2 dr = 10^{11}
M_{\astrosun}\,.
\end{equation}

Alternatively, as a realistic hot gas density distribution a numerical
model of Feldmann, Hooper \& Gnedin~\cite{Feldmann}, based on
high-resolution cosmological and hydrodynamical simulations is used. We assume that the hot gas halo contains
mainly hydrogen and helium and use the density profile obtained for ionized hydrogen distribution with the normalization
as stated above.

The electron number density is directly related to the baryonic density:
\begin{equation}
n_e = \frac{\rho}{m_p}\ \left(1-Y_P/2\right)\,\\
\end{equation}
where $m_p$ is a proton mass and $Y_P \approx 0.25$ -- helium mass fraction taken from~\cite{Planck2015}.

\section{Data set}
\label{sec:data}

In this paper we analyze pulsars in the Milky Way, in the Large
Magellanic Cloud (LMC) and in the Small Magellanic Cloud (SMC). There
are no known pulsars in Andromeda Galaxy (M31) yet, but the future
observations will likely discover ones~\cite{Hessels}.

In order to calculate the dispersion measure of a pulsar one needs to
know distance from it to the Galactic center. Usually it is calculated
from experimental DM, but there are two kind of pulsars with
independent distance in the Milky Way: pulsars with measured
parallaxes and pulsars in the globular clusters~\cite{Gaensler}.

We analyzed 144 pulsars in 28 globular clusters \cite{Globulars},
\cite{Globularsparameters} and 63 pulsars with parallaxes
\cite{Verbiest}. Within the one globular cluster analytical dispersion
measure is the same for all pulsars and experimental one differs
insignificantly. The experimental dispersion measure values were
picked up from ATNF (Australia Telescope National Facility) Pulsar
Catalogue~\cite{catalogue}. The data used for Milky Way Galaxy pulsars
data are given in Tables \ref{table:clusters} and
\ref{table:parallaxes}. The data for LMC and SMC are given in Table
\ref{table:cloudspulsars}. The galactic coordinates for M31 are $l =
    {-21.6}^{\circ}$, $b = {121.2}^{\circ}$ and distance $D =
    784.9\,\mbox{kpc}$.

\section{Results}
\label{sec:results}

\begin{figure}[h!]
\includegraphics[width=0.48\textwidth]{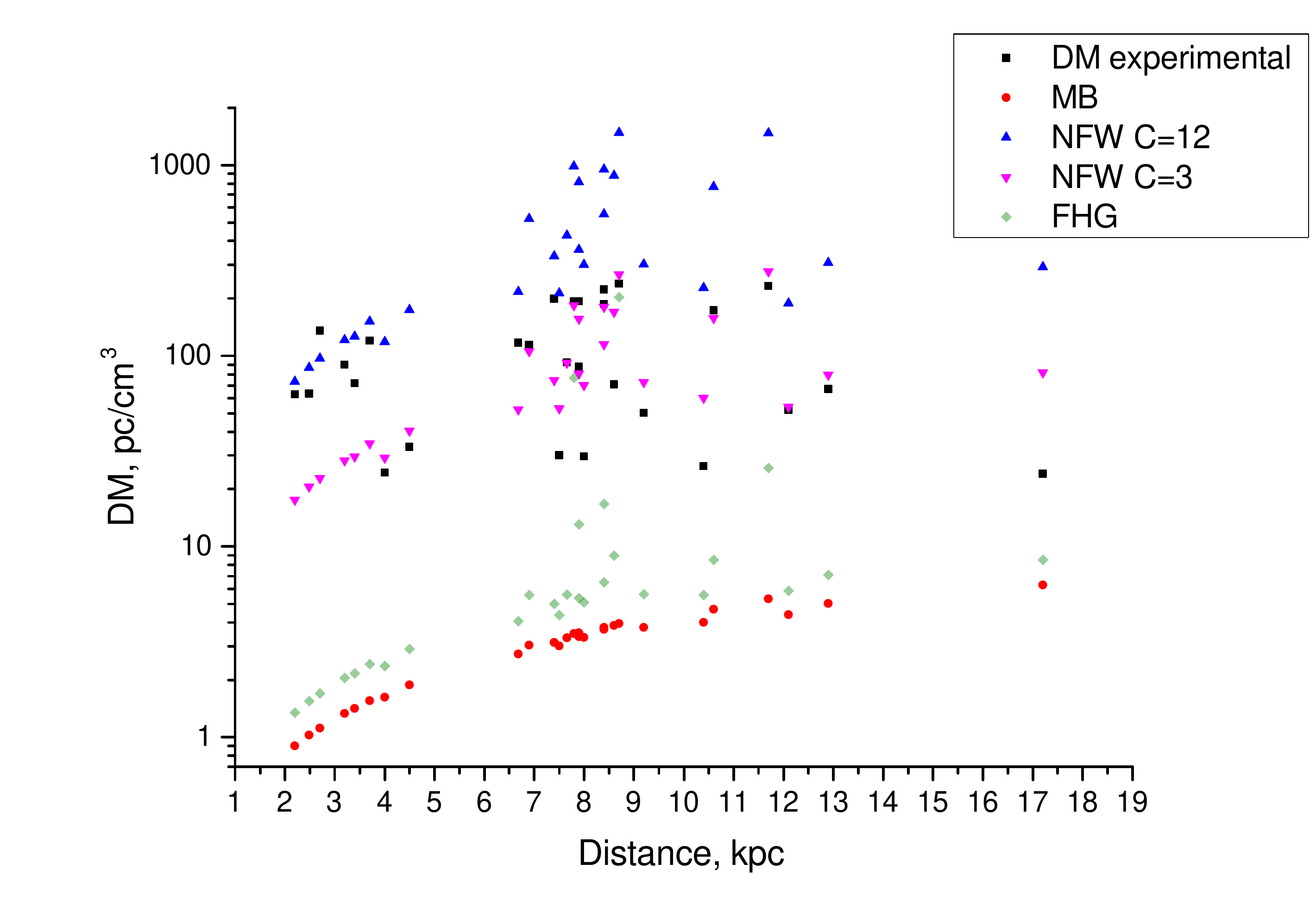}
\caption{Dispersion measure of Milky Way globular clusters'
  pulsars. ATNF data is plotted in comparison with models: MB, NFW $C
  = 12$, NFW $C = 3$ and FHG.}
\label{figure:clusters}
\end{figure}

\begin{figure}[h!]
\includegraphics[width=0.48\textwidth]{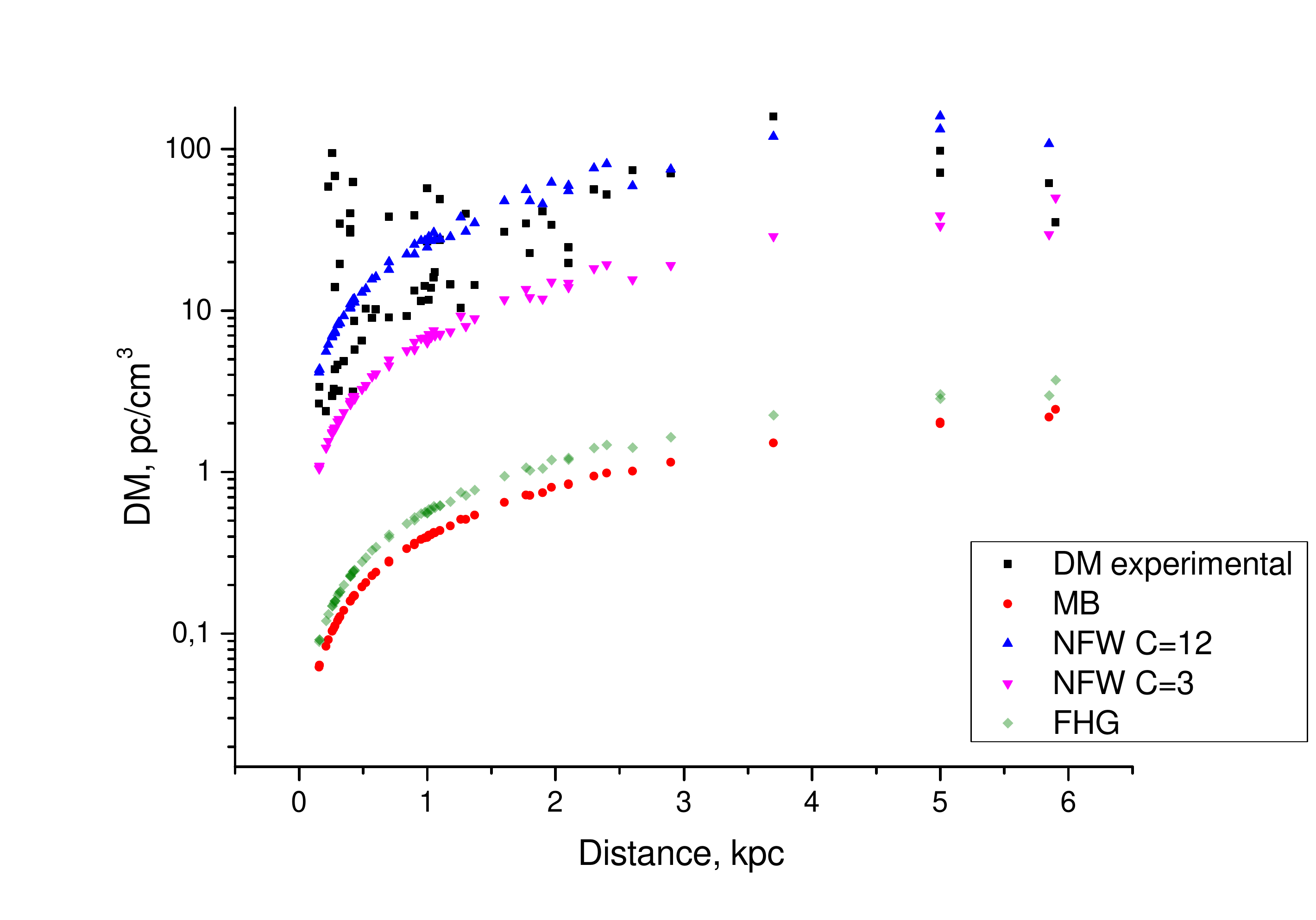}
\caption{Dispersion measure of Milky Way pulsars with parallaxes. ATNF
  data is plotted in comparison with models: MB, NFW $C = 12$, NFW $C
  = 3$ and FHG.}
\label{figure:parallaxes}
\end{figure}

The dispersion measure is calculated for each pulsar in our data set
within 4 different models: NFW model with $C=12$ and $C=3$, MB model
and FHG model. Comparisons of experimental and analytical dispersion
measures are shown in Figures \ref{figure:clusters} and
\ref{figure:parallaxes}, and in Table \ref{table:magelcloudsdm}. The
predicted DM for Andromeda is given in Table \ref{table:andromedadm}.

The dispersion measure of the particular pulsar is summed up from
  the intrinsic DM and from the effect of ISM. In this paper we
  concentrate only of the contribution of hot gas halo. Therefore, for
  a viable model the result of calculation may not be greater than the
  measured values. For the both NFW $C=3$ and $C=12$ there are pulsars
  with model DM greater than the experimental one and therefore for
  these energy density the model of ionized hot gas halo contradicts to
  the pulsar data. On the other hand, more realistic MB and FHG
models are not excluded with the method of the paper. The predicted DM
in these models do not exceed the observed one for all Milky Way, LMC
and SMC pulsars.


\section{Conclusion}
\label{sec:conclusion}

The pulsar DM data rule out NFW density model of the hot gas halo for
the both cases $C = 3$ and $C = 12$. At the same time more
realistic MB and FHG models are not constrained with this method.

In future, the studies using Square Kilometre Array (SKA) may double
or triple the known pulsar population \cite{Hessels}. For our study,
finding new pulsars on various latitudes and longitudes will provide
all-round interstellar medium probe with the higher reliability and
accuracy. Moreover, hot gas halos of the nearby massive galaxies
may be probed through the Sunyaev-Zel'dovich distortion of the cosmic
microwave background radiation~\cite{Singh}.

\section*{Acknowledgements}
The authors are indebted to Maxim Libanov, Maxim Pshirkov and Sergey
Troitsky for inspiring discussions. The work is supported by the
Russian Science Foundation grant 14-22-00161.

\begin{center}
\begin{table*}[c]
\caption{List of Milky Way globular clusters}
\smallskip
\label{table:clusters}
\begin{tabular}{|c|c|c|c|c|}
\hline
&&&&\\
\textbf{Globular cluster name} & \textbf{Galactic longitude ($deg$)} &\textbf{Galactic latitude ($deg$)} & \textbf{Distance ($kpc$)} & \textbf{DM ($cm/{pc}^3$)}\\
&&&&\\
\hline
47 Tucanae (NGC 104) & 305,9 & -44,89 & 4 & 24,4\\ \hline
NGC 1851 & 244,51 & -35,04 & 12,1 & 52,1489\\ \hline
M 53 (NGC 5024) & 322,96 & 79,69 & 17,2 & 24\\ \hline
M 3 (NGC 5272) & 42,41 & 78,71 & 10,4 & 26,4\\ \hline
M5 (NGC 5904) & 3.86 & 46.8 & 8 & 29.6\\ \hline
NGC 5986 & 337.02 & 13.27 & 7.66 & 92.17\\ \hline
M4 (NGC 6121) & 350.97 & 15.97 & 2.2 & 62.8633\\ \hline
M13 (NGC 6205) & 59.01 & 40.91 & 7.5 & 30\\ \hline
M62 (NGC 6266) & 353.57 & 7.32 & 6.9 & 114.06\\ \hline
NGC 6342 & 4.9 & 9.73 & 8.6 & 71\\ \hline
NGC 6397 & 338.17 & -11.96 & 3.4 & 71.8\\ \hline
Terzan 5 & 3.84 & 1.69 & 8.7 & 238.5\\ \hline
NGC 6440 & 7.73 & 3.8 & 8.4 & 222.8\\ \hline
NGC 6441 & 353.53 & -5.01 & 11.7 & 232.2\\ \hline
NGC 6517 & 19.23 & 6.76 & 10.6 & 172.98\\ \hline
NGC 6522 & 1.02 & -3.93 & 7.8 & 193\\ \hline
NGC 6539 & 20.8 & 6.78 & 8.4 & 186.38\\ \hline
NGC 6544 & 5.84 & -2.2 & 2.7 & 135.6\\ \hline
NGC 6624 & 2.79 & -7.91 & 7.9 & 87.61\\ \hline
M28 (NGC 6626) & 7.8 & -5.58 & 3.7 & 120.1\\ \hline
NGC 6652 & 1.53 & -11.38 & 2.49 & 63.35\\ \hline
M22 (NGC 6656) & 9.89 & -7.55 & 3.2 & 89.7\\ \hline
NGC 6749 & 36.2 & -2.2 & 7.9 & 192.846\\ \hline
NGC 6752 & 336.49 & -26.53 & 4.5 & 33.36\\ \hline
NGC 6760 & 36.11 & -3.92 & 7.4 & 199.684\\ \hline
M71 (NGC 6838) & 56.74 & -4.56 & 6.68 & 117\\ \hline
M15 (NGC 7078 & 65.01 & -27.31 & 12.9 & 66.87\\ \hline
M30 (NGC 7099) & 27.18 & -48.83 & 9.2 & 50.07\\ \hline
\end{tabular}
\end{table*}
\end{center}

\begingroup
\squeezetable
\begin{table*}[c]
\caption{List of Milky Way pulsars with parallaxes}
\smallskip
\label{table:parallaxes}
\begin{tabular}{|c|c|c|c|c|}
\hline
&&&&\\
\textbf{Pulsar name} & \textbf{Galactic longitude ($deg$)} &\textbf{Galactic latitude ($deg$)} & \textbf{Distance ($kpc$)} & \textbf{DM ($cm/{pc}^3$)}\\
&&&&\\
\hline
J0030+0451 & 113,14 & -57,61 & 0,28 & 4,333\\ \hline
J0034-0721 & 110,42 & -69,82 & 1,03 & 13,76517\\ \hline
J0108-1431 & 140,93 & -76,82 & 0,21 & 2,38\\ \hline
J0139+5814 & 129,22 & -4,04 & 2,6 & 73,779\\ \hline
J0218+4232 & 139,51 & -17,53 & 5,85 & 61,252\\ \hline
J0332+5434 & 145 & -1,22 & 1 & 26,7641\\ \hline
J0358+5413 & 148,19 & 0,81 & 1 & 57,142\\ \hline
J0437-4715 & 253,39 & -41,96 & 0,156 & 2,64476\\ \hline
J0452-1759 & 217,08 & 34,09 & 0,4 & 39,903\\ \hline
J0454+5543 & 152,62 & 7,55 & 1,18 & 14,495\\ \hline
J0538+2817 & 179,72 & -1,69 & 1,3 & 39,57\\ \hline
J0613-0200 & 210,41 & -9,3 & 0,9 & 38,77919\\ \hline
J0630-2834 & 236,95 & -16,76 & 0,32 & 34,468\\ \hline
J0659+1414 & 201,11 & 8,26 & 0,28 & 13,977\\ \hline
J0720-3125 & 244,16 & -8,16 & 0,4 & --\\ \hline
J0737–3039AB & 245,24 & -4,5 & 1,1 & 48,92\\ \hline
J0751+1807 & 202,73 & 21,09 & 0,4 & 30,2489\\ \hline
J0814+7429 & 140 & 31,62 & 0,432 & 5,733\\ \hline
J0820-1350 & 235,89 & 12,59 & 1,9 & 40,938\\ \hline
J0826+2637 & 196,96 & 31,74 & 0,32 & 19,454\\ \hline
J0835-4510 & 263,55 & -2,79 & 0,28 & 67,99\\ \hline
J0922+0638 & 225,42 & 36,39 & 1,1 & 27,271\\ \hline
J0953+0755 & 228,91 & 43,7 & 0,26 & 2,958\\ \hline
J1012+5307 & 160,35 & 50,86 & 0,7 & 9,0233\\ \hline
J1017-7156 & 291,56 & -12,55 & 0,26 & 94,22407\\ \hline
J1022+1001 & 231,79 & 51,1 & 0,52 & 10,2521\\ \hline
J1023+0038 & 243,49 & 45,78 & 1,37 & 14,325\\ \hline
J1024-0719 & 251,7 & 40,52 & 0,49 & 6,4852\\ \hline
J1045-4509 & 280,85 & 12,25 & 0,23 & 58,1662\\ \hline
J1136+1551 & 241,9 & 69,2 & 0,35 & 4,8451\\ \hline
J1239+2453 & 252,45 & 86,54 & 0,84 & 9,242\\ \hline
J1300+1240 & 311,31 & 75,41 & 0,6 & 10,1655\\ \hline
J1456-6843 & 313,87 & -8,54 & 0,43 & 8,6\\ \hline
J1509+5531 & 91,33 & 52,29 & 2,1 & 19,613\\ \hline
J1537+1155 & 19,85 & 48,34 & 1,01 & 11,61436\\ \hline
J1543+0929 & 17,81 & 45,78 & 5,9 & 35,24\\ \hline
J1559-4438 & 334,54 & 6,37 & 2,3 & 56,1\\ \hline
J1600-3053 & 344,09 & 16,45 & 2,4 & 52,3262\\ \hline
J1614-2230 & 352,64 & 20,19 & 1,77 & 34,4865\\ \hline
J1643-1224 & 5,67 & 21,22 & 0,42 & 62,4121\\ \hline
J1713+0747 & 28,75 & 25,22 & 1,05 & 15,9915\\ \hline
J1738+0333 & 27,72 & 17,74 & 1,97 & 33,778\\ \hline
J1744-1134 & 14,79 & 9,18 & 0,42 &  3,13908\\ \hline
J1856-3754 & 358,61 & -17,21 & 0,16 & --\\ \hline
J1853+1303 & 44,87 & 5,37 & 1,6 & 30,5701\\ \hline
J1857+0943 & 42,29 & 3,06 & 0,9 & 13,3\\ \hline
J1900-2600 & 10,34 & -13,45 & 0,7 & 37,994\\ \hline
J1909-3744 & 359,73 & -19,6 & 1,26 & 10,3934\\ \hline
J1932+1059 & 47,38 & -3,88 & 0,31 & 3,18\\ \hline
J1935+1616 & 52,44 & -2,09 & 3,7 & 158,521\\ \hline
J1939+2134 & 57,51 & -0,29 & 5 & 71,0398\\ \hline
J2018+2839 & 68,1 & -3,98 & 0,98 & 14,172\\ \hline
J2022+2854 & 68,86 & -4,67 & 2,1 & 24,64\\ \hline
J2022+5154 & 87,86 & 8,38 & 1,8 & 22,648\\ \hline
J2048-1616 & 30,51 & -33,08 & 0,95 & 11,456\\ \hline
J2055+3630 & 79,13 & -5,59 & 5 & 97,314\\ \hline
J2124-3358 & 10,93 & -45,44 & 0,3 & 4,601\\ \hline
J2129-5721 & 338,01 & -43,57 & 0,4 & 31,853\\ \hline
J2144-3933 & 2,79 & -49,47 & 0,16 & 3,35\\ \hline
J2145-0750 & 47,78 & -42,08 & 0,57 & 8,9977\\ \hline
J2157+4017 & 90,49 & -11,34 & 2,9 & 70,857\\ \hline
J2222-0137 & 62,02 & -46,08 & 0,27 & 3,27511\\ \hline
J2313+4253 & 104,41 & -16,42 & 1,06 & 17,2758\\ \hline
\end{tabular}
\end{table*}
\endgroup

\begin{center}
\begin{table*}[c]
\caption{List of of Large Magellanic Cloud and Small Magellanic Cloud pulsars}
\smallskip
\label{table:cloudspulsars}
\begin{tabular}{|c|c|c|c|c|}
\hline
&&&&\\
\textbf{Pulsar name} & \textbf{Galactic longitude ($deg$)} &\textbf{Galactic latitude ($deg$)} & \textbf{Distance ($kpc$)} & \textbf{DM ($cm/{pc}^3$)}\\
&&&&\\
\hline
J0449-7031 & 282.29 & -35.51 & 53.7 & 65.83\\ \hline
J0451-67 & 278.41 & -36.29 & 53.7 & 45\\ \hline
J0455-6951 & 281.29 & -35.19 & 53.7 & 94.89\\ \hline
J0456-7031 & 282.05 & -34.97 & 53.7 & 100.3\\ \hline
J0502-6617 & 276.87 & -35.3 & 53.7 & 68.9\\ \hline
J0519-6932 & 280.29 & -33.25 & 53.7 & 119.4\\ \hline
J0522-6847 & 279.35 & -33.17 & 53.7 & 126.45\\ \hline
J0529-6652 & 276.97 & -32.76 & 53.7 & 103.2\\ \hline
J0532-6639 & 276.67 & -32.48 & 53.7 & 69.3\\ \hline
J0534-6703 & 277.13 & -32.28 & 53.7 & 94.7\\ \hline
J0535-6935 & 280.08 & -31.94 & 53.7 & 93.7\\ \hline
J0537-6910 & 279.56 & -31.74 & 53.7 & --\\ \hline
J0540-6919 & 279.72 & -31.52 & 53.7 & 146.5\\ \hline
J0543-6851 & 279.13 & -31.24 & 53.7 & 131\\ \hline
J0555-7056 & 281.46 & -30.12 & 53.7 & 73.4\\ \hline
J0045-7042 & 303.65 & -46.42 & 62.4 & 70\\ \hline
J0111-7131 & 300.67 & -45.51 & 62.4 & 76\\ \hline
J0113-7220 & 300.62 & -44.69 & 62.4 & 125.49\\ \hline
J0131-7310 & 298.94 & -43.65 & 62.4 & 205.2\\ \hline
J0045-7319 & 303.51 & -43.8 & 62.4 & 105.4\\ \hline
\end{tabular}
\end{table*}
\end{center}

\begin{center}
\begin{table*}[c]
\caption{Dispersion measure for LMC and SMC pulsars}
\smallskip
\label{table:magelcloudsdm}
\begin{tabular}{|c|c|c|c|c|c|}
\hline
&&&&&\\
\textbf{Pulsar name} & \textbf{Experimental DM} & \textbf{MB} & \textbf{NFW C=12} & \textbf{NFW C=3} & \textbf{FHG}\\
&&&&&\\
\hline
J0449-7031 & 65.83 & 14.85 & 381.1 & 123.4 & 20.0\\ \hline
J0451-67 & 45 & 14.7 & 366.7 & 120.0 & 19.7\\ \hline
J0455-6951 & 94.89 & 14.8 & 377.5 & 122.5 & 19.8\\ \hline
J0456-7031 & 100.3 & 14.8 & 380.4 & 123.2 & 19.9\\ \hline
J0502-6617 & 68.9 & 14.7 & 361.8 & 118.8 & 19.7\\ \hline
J0519-6932 & 119.4 & 14.8 & 374.7 & 121.9 & 19.8\\ \hline
J0522-6847 & 126.45 & 14.8 & 371.3 & 121.1 & 19.8\\ \hline
J0529-6652 & 103.2 & 14.7 & 362.8 & 119.1 & 19.7\\ \hline
J0532-6639 & 69.3 & 14.7 & 361.85 & 118.8 & 19.7\\ \hline
J0534-6703 & 94.7 & 14.7 & 363.5 & 119.2 & 19.7\\ \hline
J0535-6935 & 93.7 & 14.8 & 374.5 & 121.8 & 19.8\\ \hline
J0537-6910 & -- & 14.8 & 372.6 & 121.4 & 19.8\\ \hline
J0540-6919 & 146.5 & 14.8 & 373.3 & 121.5 & 19.8\\ \hline
J0543-6851 & 131 & 14.8 & 371.2 & 121.0 & 19.8\\ \hline
J0555-7056 & 73.4 & 14.85 & 380.65 & 123.3 & 19.9\\ \hline
J0045-7042 & 70 & 16.8 & 454.4 & 142.65 & 22.8\\ \hline
J0111-7131 & 76 & 16.7 & 444.8 & 140.5 & 22.7\\ \hline
J0113-7220 & 125.49 & 16.7 & 446.6 & 140.9 & 22.75\\ \hline
J0131-7310 & 205.2 & 16.7 & 442.0 & 139.9 & 22.7\\ \hline
J0045-7319 & 105.4 & 16.8 & 461.1 & 144.15 & 22.9\\ \hline
\end{tabular}
\end{table*}
\end{center}

\begin{center}
\begin{table*}[c]
\caption{Model predictions for Andromeda galaxy DM}
\smallskip
\label{table:andromedadm}
\begin{tabular}{|c|c|c|c|}
\hline
&&&\\
\textbf{MB} & \textbf{NFW C=12} & \textbf{NFW C=3} & \textbf{FHG}\\
&&&\\
\hline
28.8 & 282.5 & 114.6 & 37.3\\ \hline
\end{tabular}
\end{table*}
\end{center}

%\begin{figure}
%\includegraphics[width=0.48\textwidth]{}
%\caption{}
%\label{histo}
%\end{figure}

\end{document}